# Formation of Ultra-High-Resistance Au/Ti/*p*-GaN Junctions and the Applications in AlGaN/GaN HEMTs

Guangnan Zhou[1,2,3,5], Yang Jiang[1,2,3], Gaiying Yang[4], Qing Wang[1,2,3], Mengya Fan[1,2,3], Lingli Jiang[1,2,3], Hongyu Yu[1,2,3,a)] and Guangrui Xia[5,a)]

[1] *School of Microelectronics, Southern University of Science and Technology (SUSTech), Shenzhen, 518055, China*

[2] *GaN Device Engineering Technology Research Center of Guangdong, SUSTech*

[3] *Key Laboratory of the Third Generation Semiconductors, SUSTech*

[4] *School of Innovation & Entrepreneurship, SUSTech*

[5] *Department of Materials Engineering, the Univerisity of British Columbia (UBC), Vancouver, V6T 1Z4, Canada*

We report a dramatic current reduction, or a resistance increase, by a few orders of magnitude of two common-anode Au/Ti/*p*-GaN Schottky junctions annealed within a certain annealing condition window (600 - 700 °C, 1 - 4 min). Results from similar common-anode Schottky junctions made of Au/*p*-GaN, Al/Ti/*p*-GaN and Au/Ti/graphene/*p*-GaN junctions demonstrated that all the three layers (Au, Ti and *p*-GaN) are essential for the increased resistance. Raman characterization of the *p*-GaN showed a decrease of the Mg-N bonding, i.e., the deactivation of Mg, which is consistent with the Hall measurement results. Moreover, this high-resistance junction structure was employed in *p*-GaN gate AlGaN/GaN HEMTs. It was shown to be an effective gate technology that was capable to boost the gate breakdown voltage from 9.9 V to 13.8 V with a negligible effect on the threshold voltage or the sub-threshold slope.

Gallium nitride (GaN) possesses many excellent physical properties, including a wide bandgap of 3.4 eV, a large critical breakdown field of 3.3 MV/cm, a high electron saturation velocity of $2.5\times10^7$ cm/s, and a good thermal conductivity up to 1.5 W/cm·K.[1] Owing to these superior properties, GaN-based devices have been very promising for a wide range of semiconductor device applications, including UV detectors,[2,3] radio frequency (RF)/microwave electronics,[4,5] gas sensors,[6,7] and high speed and high-power electronics.[8] Besides, GaN-based light-emitting diodes (LEDs) and laser diodes (LDs) have already been commercialized for a variety of lighting and display applications.[10]

Metal/*p*-GaN junctions are essential in many of these electronic devices. For example, low-resistance and transparent ohmic metal contacts to *p*-GaN are crucial to improve current injection and light extraction efficiency of GaN-based LEDs. For *p*-channel GaN-based transistors, thermally stable and low-resistance ohmic metal/*p*-GaN contacts are highly desired, where

___________________________

a) Authors to whom correspondence should be addressed: yuhy@sustech.edu.cn; guangrui.xia@ubc.ca.

Ni- metals are most widely used for its high work function.[10,11] Meanwhile, for high electron mobility transistors (HEMTs) with *p*-GaN gate, Schottky type metal/*p*-GaN gate with high Schottky barrier height is preferred to reduce the gate leakage current and increase the threshold voltage. Ti- or TiN- based metal electrodes are usually adopted due to their low work functions.[12,13] Even though Ti/Al or Ti/Au metal stacks have been widely used as the *p*-GaN gate metal,[13,14] the impact of metal/*p*-GaN junction on the HEMTs' performance parameters such as the threshold voltage and the gate breakdown voltage has seldom been investigated. Only a handful of recent papers discussed the influence of the different metals and fabrication processing conditions on the HEMTs performances. G. Greco *et al.* demonstrated that the HEMTs with Ti/Al gate showed a considerable high leakage current when subjected to 800 °C 1 min annealing.[14] The importance of the metal contacts on *p*-GaN still deserves further investigation to optimize the performances of these electronic devices.

In this letter, we present a detailed investigation of Au/Ti/*p*-GaN junctions, focusing on the impact of annealing thermal budget on the resistance of these junctions. For the first time, we report that ultra-low current, i.e., ultra-high resistance, can be achieved using Au/Ti/*p*-GaN within a particular process window, which is contrary to the observations in Ref. [14,15]. Au/*p*-GaN, Al/Ti/*p*-GaN and Au/Ti/graphene/*p*-GaN junctions were also studied to investigate the current reduction mechanism. The Au/Ti/*p*-GaN junction were also been characterized by cross-sectional scanning-transmission electron microscopy (STEM) and Raman spectroscopy. Moreover, this novel Au/Ti/*p*-GaN scheme has been used in *p*-GaN gate HEMTs to investigate its benefits in device performance. Compared to the conventional metal gate without annealing, the devices with optimized Au/Ti/*p*-GaN gate have shown superior gate breakdown voltages with little changes on the threshold voltages or the sub-threshold slopes.

The 100 nm *p*-GaN/15 nm AlGaN/0.7 nm AlN/GaN buffer epitaxial structures used in the experiments were grown on (111) Si substrates by metal-organic chemical vapor deposition (MOCVD) provided by Enkris Semiconductor Inc. The *p*-GaN layers were doped with Mg to a concentration of $4 \times 10^{19}$ cm$^{-3}$. As seen in FIG. 1, the test structure consists of two common-anode Schottky junctions in series at a distance of 20 μm, with one forward-biased and the other reverse-biased.[16,17] Prior to the metal deposition, the *p*-GaN surface were thoroughly cleaned using acetone, ethanol, 1:4 HCl:H$_2$O, and then rinsed in DI water each for 10 minutes. Au/Ti (100nm/40 nm) layers were deposited by the e-beam evaporator and were patterned by a lift-off technique. The metals featured a width of 100 μm and a length of 25 μm. To investigate the annealing effect, the samples were annealed at 400, 500, 600, 700 and 800 °C respectively for 1 min in an N$_2$ ambient using Rapid Thermal Process (RTP) by Annealsys AS-one system.

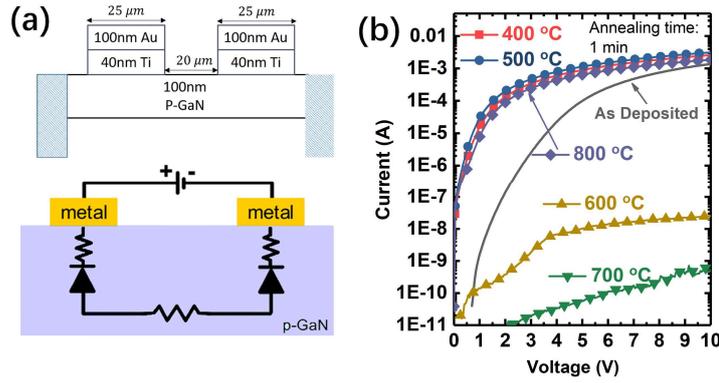

FIG. 1. (a) Schematic of the investigated common-anode Au/Ti/*p*-GaN Schottky contacts, the measurement setup and its equivalent circuit. (b) I–V characteristics of the common-anode Au/Ti/*p*-GaN Schottky contacts with different thermal annealing conditions: the plain solid line is from the as-deposited sample, and the solid lines with symbols are from the samples annealed at 400 to 800 °C for 1 min.

Current-voltage (I–V) measurements were carried out in the bias range of -10 to 10 V. FIG. 1(b) shows the I-V characteristics of the two common-anode junctions with and without annealing. The data from -10 to 0 V are not presented due to the symmetry of the I-V curves. As the reverse-biased junction has a much larger resistance than the forward-biased junction, the I-V behavior is determined by the reverse-biased junction, which is the junction on the left in Fig.1(a). Annealing commonly leads to a higher current due to the lower Schottky barrier height.[14,15] Indeed, the I–V measurements showed an increase in the current when the annealing temperature was 400 °C or 500 °C, which was as expected. However, when the temperature reached 600 °C, the current decreased significantly by approximately five orders of magnitude. If the annealing temperature further increased to 700 °C, the current would further drop (< $10^{-9}$ A). More interestingly, when the annealing temperature was 800 °C, the current increased to a level comparable to the case of 500 °C annealing. In summary, the resistance of this metal/*p*-GaN junction was found to have a very unusual dependence on the annealing temperature, which was first reported in literature to the best of our knowledge.

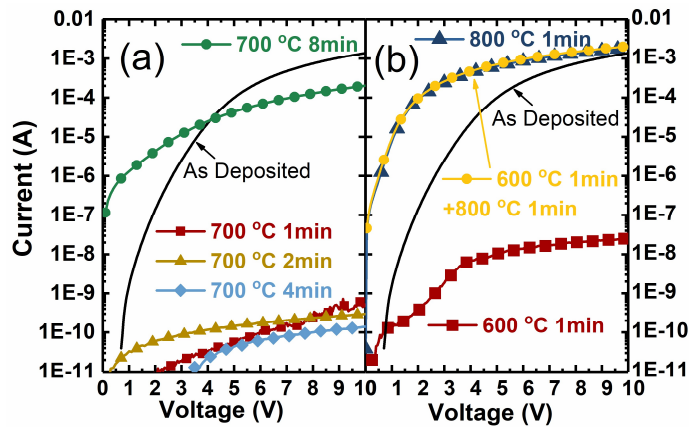

FIG. 2. I–V characteristics in Au/Ti/*p*-GaN Schottky contacts, (a) for the samples annealed at 700 °C from 1 min to 8 min; (b) for the sample annealed at 600 °C for 1min followed by 800 °C 1 min annealing.

To further investigate the current dependence on the annealing thermal budget, some samples were annealed for 1-8 min at T = 700 °C. As shown in FIG. 2, the current remained considerably low when the annealing time ranged from 1-4 min, whereas it was back to Schottky-like behavior when the annealing time was 8 min. Furthermore, for the sample annealed at 600 °C for 1 min, the current surged significantly after further annealing at 800 °C for 1 min, which was finally equivalent to the case of 800 °C 1 min annealing only. These data demonstrate that the I-V characteristics of the Au/Ti/*p*-GaN Schottky contacts are highly dependent on the annealing temperature and time. Within a certain process window, the Au/Ti/*p*-GaN contacts can be very high resistive instead of Schottky-like or ohmic. Specifically, after 700 °C 1-4 min annealing, the current across the Schottky contacts is around six orders of magnitude lower than that in as-deposited contacts at V = 5V.

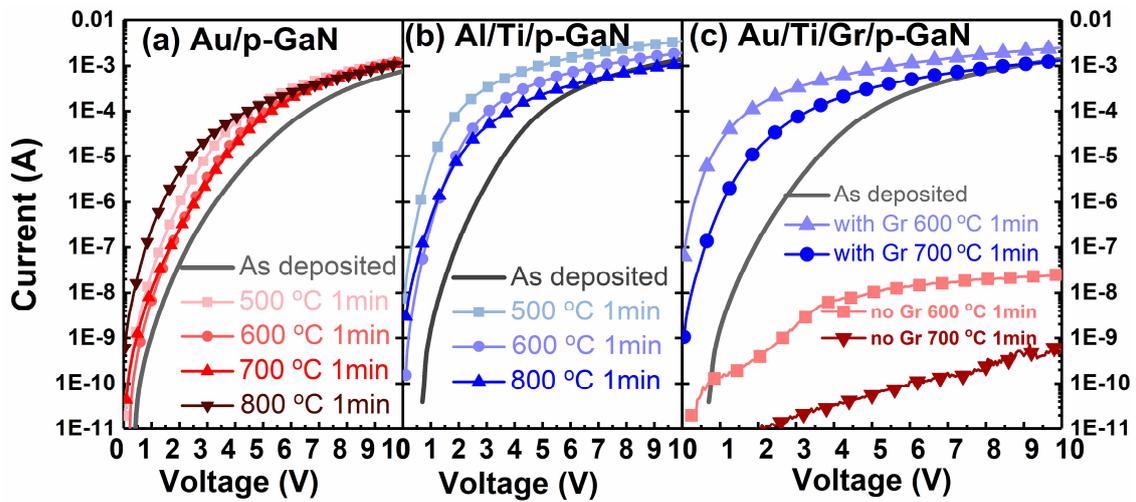

FIG. 3. I–V characteristics of the common-anode *p*-GaN Schottky contacts with different metals, (a) Au/*p*-GaN; (b) Al/Ti/*p*-GaN; (c) Au/Ti/graphene/*p*-GaN (Au/Ti/Gr/*p*-GaN) compared with Au/Ti/*p*-GaN.

Common-anode diodes with Au/*p*-GaN (no Ti) and Al/Ti/*p*-GaN (no Au) contacts were also fabricated and annealed at 500 to 800 °C for 1 min for comparison. Current reduction after annealing was not observed in either case, as shown in FIG. 3(a) and 3(b). The testing results are consistent with literature data that the annealing treatments will result in a higher current due to a lower metal/*p*-GaN barrier. Thus, we can deduce that both Ti and Au are necessary to form the high-resistance junction. To assess the role of the *p*-GaN in this process, graphene was inserted between Ti and *p*-GaN as a diffusion barrier. The strong sp2 hexagonal bonding of graphene can act as a strong barrier to atom diffusion and has been utilized as a diffusion barrier in many applications.[18-21] FIG. 3(c) compares the contact of Au/Ti/graphene/*p*-GaN (Au/Ti/Gr/*p*-GaN) and Au/Ti/*p*-GaN with and without annealing. The current in the as-deposited samples with and without graphene were comparable. After annealing at 600 °C or 700 °C for 1 min, the current in the contact with graphene further increased, contrary to the sample without graphene. The fact that the graphene barrier layer between the metals and *p*-GaN can prevent the current decrease demonstrates that *p*-GaN is also essential to form the high-resistance junction.

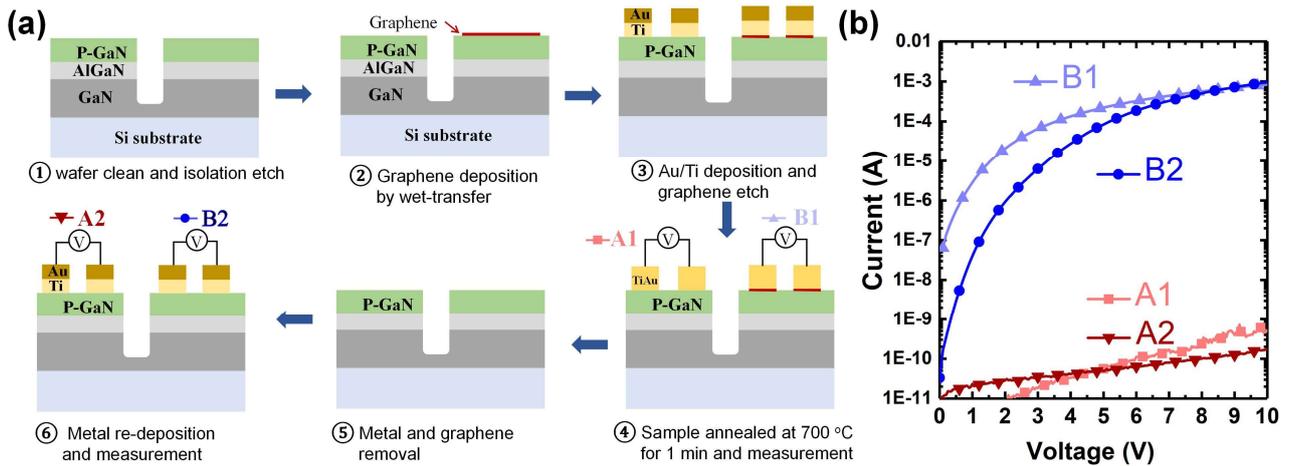

FIG. 4. (a) Schematic of the process flow of the metal re-deposition experiment; (b) I–V characteristics of the Au/Ti/(graphene)/p-GaN Schottky contacts with metal re-deposition.

To identify the critical location and analyze the formation mechanism of the high-resistance junction, a metal re-deposition experiment was designed and conducted as illustrated in FIG. 4(a). The idea was to measure I-V of the diodes with the original annealed Au/Ti/$p$-GaN (A1), remove the metal layers, redeposit new Au/Ti metals, and then measure the I-V of the new Au/Ti/$p$-GaN common-anode diodes (A2). This should reveal the role of the Au/Ti metals. Original Au/Ti/Gr/$p$-GaN diodes (B1) and new Au/Ti/p-GaN diodes (B2) were also fabricated on the same wafer and measured for comparison as graphene served as a barrier layer between Au/Ti and p-GaN.

Single-layer graphene grown by chemical vapor deposition (CVD) on Cu foils was transferred to part of the $p$-GaN sample surface via the "polymethyl methacrylate (PMMA)-mediated" wet-transfer approach.[20,22] After the deposition and patterning of 100 nm Au/40 nm Ti, graphene was etched by $O_2$ plasma using the patterned metals as the self-aligned mask. The sample was annealed at 700 °C for 1 min and then I-V measurement was performed on the common-anode diodes without and with graphene (A1 and B1). To completely remove the metals and graphene, the sample was first cleaned by acid, then exposed to $O_2$ plasma to remove the graphene. Before the re-deposition of Au/Ti, the sample was again cleaned with acid to remove the oxidation layer or any metal residuals. As shown in FIG. 4(b), the I-V characteristics of the common-anode diodes with the new Au/Ti layers (A2 and B2) were similar to their counterparts with the original Au/Ti (A1 and B1), which indicated that the metal layers are not the key regions to the drastic resistance change. Instead, the high-resistance regions should locate in the $p$-GaN layers that were annealed in contact with Au/Ti without the graphene barriers. A possible explanation is that the Au/Ti layers can deactivate the Mg in GaN by forming an electrically inactive complex with Mg within this annealing window. Under higher temperature or longer annealing time, the Mg dopant can be re-activated.

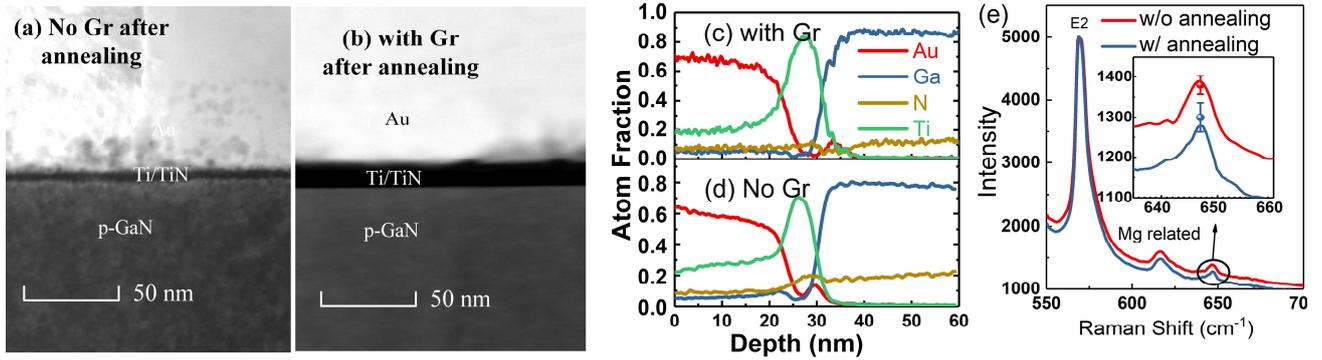

FIG. 5. Cross-section STEM of the samples after 700 °C 1 min annealing in $N_2$: (a) Au/Ti/*p*-GaN and (b) Au/Ti/graphene/*p*-GaN. (c) EDS of the Au/Ti/graphene/*p*-GaN after annealing; (d) EDS of the Au/Ti/*p*-GaN after annealing; (e) Raman spectra of the *p*-GaN with/without Au/Ti annealing; Inset: zoom-in of the Mg-related peak.

To get further insights into the Au/Ti/*p*-GaN junctions, STEM images were obtained for the Au/Ti/*p*-GaN and Au/Ti/graphene/*p*-GaN diodes with 700 °C 1 min annealing. As shown in FIG. 5(a) and 5(b), Ti has diffused into Au. Besides, TiN has formed between the Ti/*p*-GaN interface confirmed by energy-dispersive spectroscopy (EDS). However, significant morphology difference that may explain the current reduction phenomenon were not found in the STEM. The samples with and without graphene after annealing have no significant differences according to the EDS analysis.

Hall measurements were performed on the Au/Ti/*p*-GaN samples with 700 °C 1 min annealing and without annealing using the van der Pauw method. It was shown that the hole concentration has reduced from $8\times10^{17}$ cm$^{-3}$ to below $1\times10^{15}$ cm$^{-3}$ in the *p*-GaN sample after annealing. Thus, it's reasonable to attribute the current reduction to the hole concentration reduction in the *p*-GaN due to the annealing in contact with the Au/Ti layers. Raman microprobe spectra were observed at room temperature using a 532 nm laser for excitation. FIG. 5(e) presents the Raman spectra of the *p*-GaN sample with and without Au/Ti annealing. For the *p*-GaN sample with Au/Ti annealing, the metals have been completely removed by acid before the Raman characterization to avoid the reflection of the laser. The sharp peak at 569.2 cm$^{-1}$ is ascribed to GaN E2 peak; and the peak at 646 cm$^{-1}$ is ascribed to the local vibration mode of Mg-N bonding.[23,24] The spectra are normalized by the peak intensity of the E2 (569.2 cm$^{-1}$) mode and shifted vertically for comparison. It's clear shown that the Mg-N mode peak has decreased in the sample with Au/Ti annealing. Raman characterization were performed at five different positions to acquire statistical data. As illustrated in the inset of FIG. 4(e), the peak intensity of the Mg-N mode in the sample with Au/Ti annealing is 5.7% lower than the sample without annealing on average. This suggests that Au/Ti may have formed complexes with Mg in *p*-GaN after annealing, which lowers the hole concentration and makes the *p*-GaN layer highly resistive.

The technology of achieving high-resistance metal/p-GaN junctions conveniently by thermal annealing and with high-repeatability is very promising for applications in GaN-based devices consisting of metal/*p*-GaN Schottky contacts (e.g., *p*-

GaN gate HEMTs, ultraviolet Schottky barrier photodetectors). To demonstrate the application of this technology, we fabricated HEMTs with Au/Ti/*p*-GaN as the gate structure.

The *p*-GaN gate HEMTs were fabricated on 100 nm *p*-GaN/15 nm Al$_{0.2}$Ga$_{0.8}$N/0.7 nm AlN/4.5 μm GaN/Si epi-structures. FIG. 6(a) shows the schematic cross-section of the devices. The fabrication flow started with *p*-GaN gate definition by a Cl-based plasma etch, followed by a Cl$_2$/BCl$_3$ plasma etch to form mesas and isolate the devices. After the deposition of Al$_2$O$_3$ as the passivation layer by atomic layer deposition (ALD), the source/drain (S/D) Ohmic contacts were formed by Ti/Al/Ti/Au (20/110/40/50 nm) deposition and annealing. After the gate window opening and the deposition of gate metal (150 nm Au/40 nm Ti), the sample was divided into two groups. One group of HEMTs were annealed at 700 °C in N$_2$ for 1 min, and the other group didn't have any gate metal annealing.

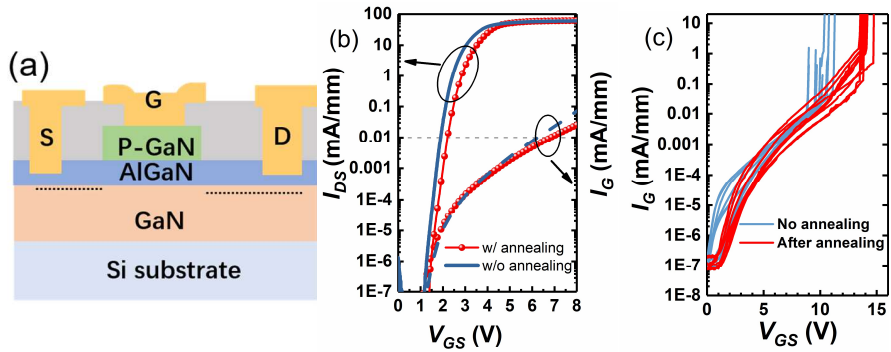

FIG. 6. (a) Cross-sectional schematic of the *p*-GaN gate HEMTs structure; (b) transfer characteristic of the HEMTs with or without gate metal annealing; (c) gate leakage characteristic and gate breakdown voltage of the HEMTs with or without gate metal annealing on ten devices for each.

The transfer characteristics of the HEMTs with or without Au/Ti gate contacts annealing are shown in FIG. 6(b). On the one hand, the annealing process step has little impact on the transfer characteristics and gate leakage characteristics of the HEMTs. Since the active P-doping concentration is only reduced in the region close to the Schottky interface with the metal, it has a limiting impact on threshold voltage and gate leakage, which are dominated by *p*-GaN/AlGaN/GaN junction.[25,26] On the other hand, a clear impact of the annealing process can be observed on the gate breakdown voltages as illustrated in FIG. 6(c). When a positive bias is applied on the gate, and the Schottky metal/*p*-GaN junction is reverse biased. The reduction of the active doping concentration close to the Schottky junction promotes a wider depletion region. Thus, the maximum electric field is lower after the annealing process, resulting in an increase of the average of gate breakdown voltage from 9.9 to 13.8 V. A similar effect has been observed in Ref. [25], where a Mg compensation process has been adopted to reduce the hole concentration close to the Schottky interface. It was shown that the reduction of hole in this region can influence the gate breakdown voltage without altering the threshold voltage, trans-conductance and sub-threshold slope. This further confirms our assumption that the high-resistance junction results from the formation of Au/Ti-Mg complexes during the annealing process, which reduces the hole concentration in $p$-GaN layer close to the metal.

In summary, we observed a significant leakage current reduction in Au/Ti/$p$-GaN Schottky diodes within a specific annealing window (600 - 700 °C, 1 - 4 min), which has not been reported for metal/$p$-GaN junctions. By comparing the Au/$p$-GaN, Al/Ti/$p$-GaN and Au/Ti/graphene/$p$-GaN Schottky diodes, we inferred that all the three layers (Au, Ti and $p$-GaN) are essential for the formation of this high-resistance junction. Also, the high-resistance regions were formed in the $p$-GaN layer instead of the metal layers. The mechanisms of the Au/Ti/$p$-GaN high-resistance junction formation were further investigated by STEM, Hall and Raman measurements. It was concluded that the increased resistance resulted from the decrease of activated Mg in $p$-GaN due to the Au/Ti annealing. This high-resistance junction structure has been employed in $p$-GaN gate HEMTs. The gate breakdown voltage was boosted from 9.9 to 13.8 V with negligible influence on the sub-threshold slope and threshold voltage. The finding of this high-resistance metal/$p$-GaN is promising for the GaN-based applications consisting of metal/$p$-GaN contacts.

This work was supported by Grant #2019B010128001 and #2017A050506002 from Guangdong Science and Technology Department, Grant #JCYJ20160226192639004 and #JCYJ20170412153356899 from Shenzhen Municipal Council of Science and Innovation. The work was conducted at SUSTech Core Research Facilities (CRF), and we would like to acknowledge the technical support from SUSTech CRF.

The data that support the findings of this study are available from the corresponding author upon reasonable request.